\documentclass[aps, prx, reprint, superscriptaddress, longbibliography, floatfix]{revtex4-2}
\usepackage{graphicx}% Include figure files
\usepackage{dcolumn}% Align table columns on decimal point
\usepackage{bm}% bold math
\usepackage{physics}
\usepackage{calrsfs}
\DeclareMathAlphabet{\pazocal}{OMS}{zplm}{m}{n}
\usepackage{amsmath}
\usepackage{amsfonts}
\usepackage{mathrsfs}
\usepackage{subfigure}
\usepackage{color}
\usepackage{xcolor}
\usepackage[colorlinks=true,linkcolor=blue,anchorcolor=red,citecolor=blue, urlcolor=blue]{hyperref}

\begin{document}
\preprint{APS/123-QED}
 
%\title{Pattern formation in quantum quenches of charge density waves}
\title{Pattern formation in charge density wave states after a quantum quench}

\author {Lingyu Yang}
\affiliation{Department of Physics, University of Virginia, Charlottesville, Virginia, 22904, USA}

\author {Yang Yang}
\affiliation{Department of Physics, University of Virginia, Charlottesville, Virginia, 22904, USA}

\author {Gia-Wei Chern}
\affiliation{Department of Physics, University of Virginia, Charlottesville, Virginia, 22904, USA}

\begin{abstract}
We study post-quench dynamics of charge-density-wave (CDW) order in the square-lattice $t$-$V$ model. The ground state of this system at half-filling is characterized by a checkerboard modulation of particle density. A generalized self-consistent mean-field method, based on the time-dependent variational principle, is employed to describe the dynamical evolution of the CDW states. Assuming a homogeneous CDW order throughout the quench process, the time-dependent mean-field approach is reduced to the Anderson pseudospin method. Quench simulations based on the Bloch equation for pseudospins produce three canonical behaviors of order-parameter dynamics:  phase-locked persistent oscillation, Landau-damped oscillation, and dynamical vanishing of the CDW order. We further develop an efficient real-space von Neumann equation method to incorporate dynamical inhomogeneity into simulations of quantum quenches. Our large-scale simulations uncover complex pattern formations in the post-quench CDW states, especially in the strong quench regime. The emergent spatial textures are characterized by super density modulations on top of the short-period checkerboard CDW order. Our demonstration of pattern formation in quenched CDW states, described by a simple broken $Z_2$ symmetry, underscores the importance of dynamical inhomogeneity in quantum quenches of many-body systems with more complex orders.
\end{abstract}
\date{\today}
\maketitle

\section{Introduction}

\label{sec:intro}

The post-quench dynamics of quantum systems has attracted enormous attention in recent years~\cite{Eisert_2015,Mitra_2018,Polkovnikov_2011,Eisert2015}. The interest is partly spurred by tremendous experimental advances both in cold atom systems~\cite{Bloch2008, Cazalilla2011} and in ultrafast techniques such as pump-probe spectroscopy~\cite{fausti2011, Matsunaga2012, Matsunaga2013, mansart2013, matsunaga2014, mitrano2016}. In particular, for cold-atomic gases trapped in an optical lattice, the parameters of the Hamiltonian, such as interaction strength between particles and lattice parameters can be tuned rapidly in time, thus providing a near-ideal platform for studying quench-induced nonequilibrium quantum behaviors. Moreover, as cold-atom systems can be well isolated from the environment, their post-quench dynamics is well approximated by a unitary evolution. A central question is whether thermalization can be reached in such closed-system quantum evolution and, if yes, what are the mechanisms and time-scales~\cite{Kollath_2007,Berges_2008,Cramer2008, Flesch2008, Rigol_2009, Dziarmaga2010, Gogolin2011, Erne2018,Kinoshita2006, Trotzky2012, schreiber2015,Mallayya_2018,Rigol_2016}. Moreover, several nonequilibrium phases and dynamical phase transitions, some of which  have no counterpart in equilibrium systems, have been demonstrated in the long-lasting prethermal states after a quantum quench~\cite{Barnett_2011,Gring_2012,Langen_2013,Marcuzzi_2013,Mitra_2013,Nessi_2014,Langen_2015,Babadi_2015,Bertini_2015,Langen_2016,Buchhold_2016}.
Understanding the nature of such prethermal states is an ongoing active research. 

For post-quench dynamics with spontaneous symmetry breaking phases, the interplay between the collective modes, as represented by order-parameter fields, and quasiparticle degrees of freedom dominates the nonequilibrium behaviors of the prethermal states~\cite{Marino_2022,Heyl_2014}. Extensive theoretical studies, including several pioneering works on the interaction quench of superconductor pairing field, have revealed three dynamical phases of the time-dependent order parameter in the collisionless limit~\cite{Barankov_2004,Barankov_2006,Yuzbashyan_2006a,Yuzbashyan_2006,Gurarie_2009}. In the phase-locked regime, the dynamics of the collective modes locks with that of the quasiparticles, giving rise to a persistent oscillation of the order parameter. For intermediate quench amplitudes, the perturbed system relaxes to a state with a reduced order parameter through the Landau-damping mechanism. The energy transfer from the collective modes to quasipartiles in this regime leads to a damped oscillation of the order parameter. Finally, in the third strongly-damped regime, corresponding to a quench toward reduced interactions, the relaxation of the system is characterized by dynamically vanishing order parameters. While more complex dynamical behaviors have also been discussed, most symmetry-breaking phases, including the well-studied BCS superconductivity (SC) and spin density waves (SDW) as well as more complex order parameters, exhibit the above three major  dynamical regimes~\cite{Foster_2013,Foster_2014,Tsuji_2013,dong2015,Peronaci_2015,Dora_2016,Blinov_2017,Chou_2017,Yoon_2017,Hannibal_2018a,Hannibal_2018,Cui_2019,Scaramazza_2019,Chen_2020}.

It is worth noting that in most previous works the prethermal states induced by interaction quenches are assumed to be spatially homogeneous, including the three main dynamical phases discussed above. For example, the Anderson pseudospin approach~\cite{Anderson_1958}, which is widely used in the study of quench dynamics of BCS superconductors, precludes the possibility of spatial fluctuation or inhomogeneity. On the other hand, since thermalization mechanisms are generally local, it is likely that the post-quench states would develop spatial inhomogeneity through a process similar to the Kibble-Zurek mechanism~\cite{Kibble_1976,Zurek_1985}. For the case of interaction quench, significant amount of energy is uniformly injected into a system. The relaxation of local regions that are separated from each other by a distance greater than the coherent length would proceed independently. Such incoherent local relaxations lead to the generation of topological defects of the order parameter, giving rise to a heterogeneous post-quench state~\cite{Damski_2005,Sengupta_2008,Zurek_2005,Mathey_2010,Chandran_2012,delCampo_2018}.

Even in the collisionless coherent regime, the highly nonlinear post-quench evolution renders the system susceptible to instabilities that could lead to inhomogeneous states. For example, it is shown that quench-induced large-amplitude coherent oscillations of the SC order parameter are unstable against parametric instability and the emergence of Cooper-pair turbulence~\cite{Dzero_2009}. Spontaneous formation of inhomogeneous superconducting states following an interaction quench has indeed been demonstrated in real-space dynamical simulations based on time-dependent Hartree-Fock-Bogoliubov theory~\cite{Chern_2019,Huang_2019}. Similar scenario has also been observed in the context of Mott transitions in Hubbard models. Although the Mott metal-insulator transition is not characterized by the broken-symmetry scenario, the on-site double occupation probability serves as an effective order parameter of a Mott transition. Interestingly, the double-occupancy of a quenched Hubbard model exhibits a coherent oscillation similar to that of conventional order parameters~\cite{Eckstein_2009,Schiro_2010}. Real-space dynamical simulations of Mott phases in the coherent collisionless limit also found post-quench states with a highly inhomogeneous distribution of the double occupancy~\cite{Chern_2017}.

While dynamical inhomogeneity in quantum quenches of many-body systems remains to be systematically investigated, pattern formation phenomena are ubiquitous in nonequilibrium systems with highly nonlinear dynamics. In particular, the mechanisms and characterizations of pattern formation in classical physics, soft-matter, and biological systems have been intensively studied for decades~\cite{Nicolis77,Cross_1993,Koch_1994,Pismen06,Cross09}. Several unifying descriptions of pattern forming systems, such as the Swift-Hohenberg equation, as well as classifications of universal behaviors have also been developed. It remains to be seen whether some of the general mathematical frameworks can be applied to pattern formation induced by quantum quenches. However, the physical mechanisms of pattern-forming instabilities in quantum quenches are intrinsically different. For most classical examples, the formation of complex patterns often is driven by energy injection and the subsequent local dissipation in an open system. On the other hand, the complex textures of order-parameter field in post-quench states result from the unitary evolution of a closed quantum system.

In this paper, we demonstrate quench-induced pattern formations in a system with a broken $Z_2$ symmetry, which is perhaps one of the simplest symmetry-breaking phases, of a fermionic system. Specifically, we present extensive simulations on the interaction quench of the square-lattice $t$-$V$ model, which exhibits a checkerboard charge density wave (CDW) order at half-filling. As the checkerboard charge modulation breaks the sublattice symmetry of the square lattice, the resultant CDW order is characterized by a $Z_2$ Ising order parameter. A time-dependent mean-field approach is employed to describe the coherent dynamics of the perturbed $t$-$V$ model. Assuming homogeneous CDW order throughout the quench process, the coherent dynamics of the post-quench CDW states can be modeled by an Anderson pseudospin theory, similar to the ones used for quench dynamics of BCS superconductors. To incorporate dynamical inhomogeneities into the quench simulations, we further develop a real-space nonlinear von~Neumann equation formulation that allows for efficient large-scale simulations.

\begin{figure}[t]
\includegraphics[width=86mm]{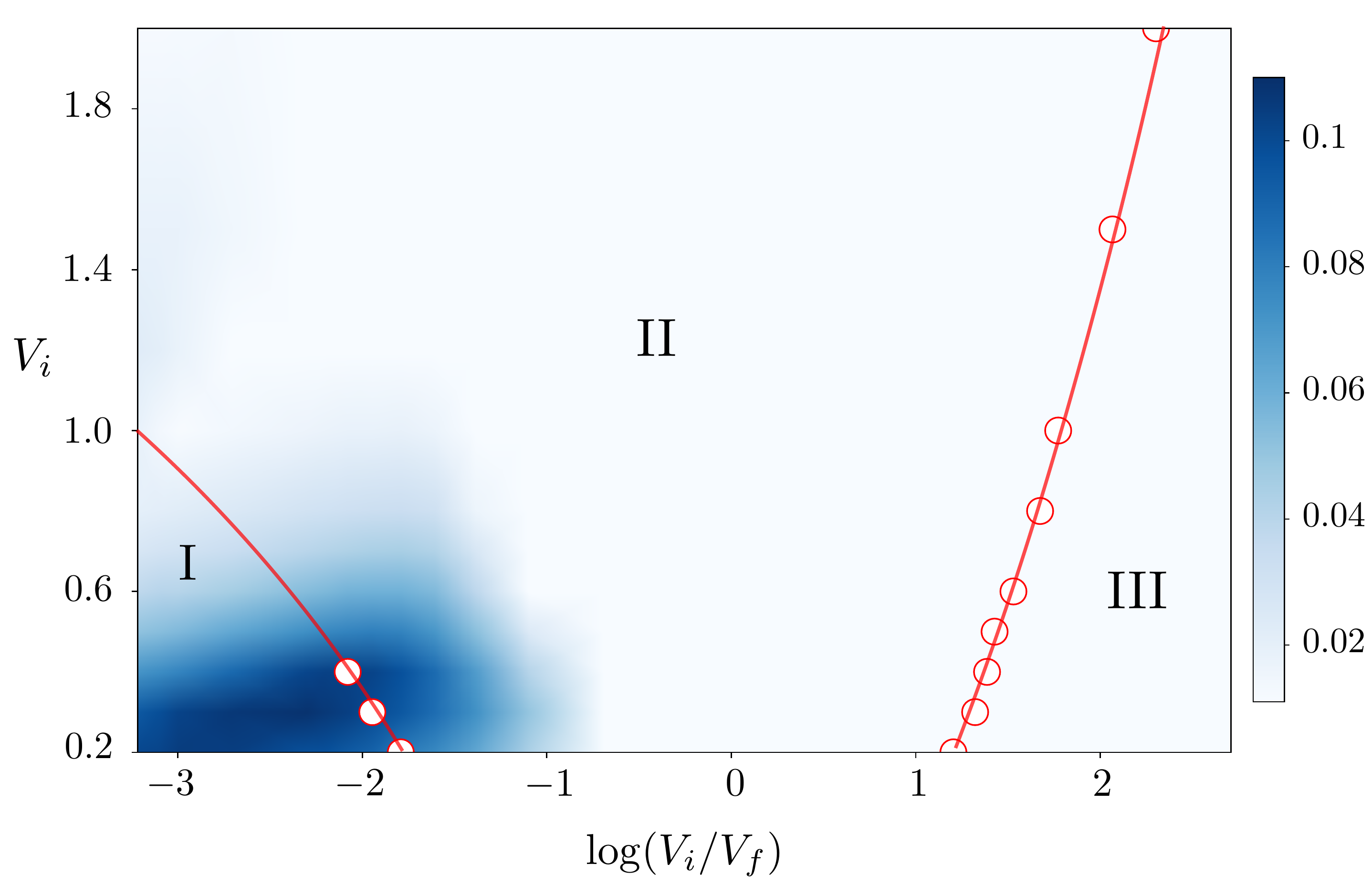}
\caption{\label{phase_diagram}
Phase diagram of post-quench CDW states of the square-lattice $t$-$V$ model. $V_i$ and $V_f$ denote the density-density repulsion before and after the sudden quench. The three dynamical phases are: (I) phase-locked persistent oscillation, (II) Landau-damped oscillation, (III) over-damped oscillation with a vanishing CDW order. The data points are obtained from the Anderson pseudospin method based on the assumption that a homogeneous CDW order persists in the post-quench states. The dashed lines are guide for eyes. Large-scale  quench simulations based on the real-space von~Neumann equation are used to study quench-induced dynamical inhomogeneity. The color intensity indicates the spatial inhomogeneity of density-modulation in the post-quench states.
}
\end{figure}

The main results of our extensive quench simulations  are summarized in a phase diagram of dynamical regimes shown in Fig.~\ref{phase_diagram}. Simulation results based on pseudospin methods show that the post-quench CDW order exhibits the three canonical dynamical phases discussed above: (I) phase-locked persistent oscillation, (II) Landau-damped oscillation with a finite asymptotic value, and (III) dynamical vanishing of the CDW order parameter. Remarkably, our real-space simulations uncover inhomogeneous CDW states with complex modulation patterns in both the phase-locked and Landau-damping regimes, suggesting an important new dynamical phase characterized by pattern formation in the post-quench states. Further characterizations of these dynamical phases will be discussed in the following sections.

The rest of the paper is organized as follows. In Sec.~\ref{sec:TDVP}, we briefly review the theory of time-dependent Hartree-Fock (TDHF) theory from the viewpoint of Dirac-Frenkel variational principle and demonstrate the approach on the $t$-$V$ model. In Sec.~\ref{sec:k-space}, an Anderson pseudospin formulation for the quench dynamics of CDW is derived from the general theory under the assumption of a homogeneous  CDW order. We then reformulate the general real-space TDHF method as a nonlinear von~Neumann equation for the single-fermion density matrix in Sec.~\ref{sec:real-space}.  The characterizations of the dynamical inhomogeneity are discussed and analyzed in detail in Sec.~\ref{sec:pattern}. Finally, we conclude the paper with a summary and outlook in Sec.~\ref{sec:conclusion}. 

% Efficient implementation of this real-space dynamical method allows us to perform large-scale quench simulations.

\section{time-dependent Hartree-Fock method for the $t$-$V$ model}

\label{sec:TDVP}

We consider the quantum quench dynamics of the spinless fermionic $t$-$V$ model on a square lattice. This model, which can be viewed as a simplified version of the Hubbard model, describes the competition between delocalization of fermions on a lattice and the nearest-neighbor density-density repulsion. Its Hamiltonian reads 
\begin{eqnarray}
\label{H_t_V}
    \hat{\mathcal{H}} = -t_{\rm nn}\sum_{\langle ij \rangle} \hat{c}^\dagger_{i} \hat{c}^{\,}_j + V \sum_{\langle ij \rangle} \hat{n}_{i} \hat{n}_j,
\end{eqnarray}
where $\hat{c}^\dagger_{i}$ ($\hat{c}^{\,}_{i}$) denotes the creation (annihilation) operator of a spinless fermion at site-$i$, $\hat{n}_{i} = \hat{c}^\dagger_{i} \hat{c}^{\,}_{i}$ is the fermion number operator. The first term above describes particle hopping between neighboring sites $\langle ij \rangle$ on a square lattice, with $t_{\rm nn}$ being the transfer coefficient. The second term with $V > 0$ represents a nearest-neighbor density-density repulsive interaction.

For a half-filled $t$-$V$ model defined on a bipartite lattice, the ground state in the strong coupling limit $V \to \infty$ is obtained when one sublattice is fully occupied, while the other sublattice is empty. This staggered arrangement of fermions breaks the $Z_2$ sublattice symmetry and can be described by an Ising order parameter. For the case that the fermionic particles are electrons, the resultant density modulation corresponds to a charge density wave (CDW) order. Although the $t$-$V$ model is also often used to describe charge-neutral fermionic cold atoms, we shall refer to a periodic particle-density modulation as a CDW for convenience. In the case of a square-lattice, a divergent Lindhard susceptibility due to a perfect Fermi surface nesting at half-filling indicates that the system is unstable against the formation of a checkerboard CDW order even in the small $V$ limit.

For a finite repulsive interaction, the $t$-$V$ model cannot be exactly solved either for equilibrium states or dynamical evolution. The 1D version of this model can be efficiently solved with high accuracy using the  density matrix renormalization group (DMRG)~\cite{White_1992,Schollwock_2005} and its time-dependent generalizations such as time-evolving block decimation (TEBD) algorithm~\cite{Vidal_2003,White_2004}. Although DMRG or tensor-network variational wave functions can also be applied to quasi-2D systems, e.g. with a cylinder geometry, such approaches are computationally more intensive with less well-controlled accuracy. Large-scale dynamical simulations of higher-dimensional systems are thus not feasible using such DMRG-based methods. 

The Hartree-Fock (HF) mean-field method proves an efficient approximation to the modeling of the emergent CDW states. The central idea of HF approximation, as in most mean-field type methods, is to reduce a many-body problem into an effective single-particle problem self-consistently. While such self-consistency can be achieved via the usual mean-field decoupling of the interaction terms, HF approximation is best understood as a variational method in which the trial many-fermion wave function has the form of a Slater determinant. The HF or its extension for superconducting pairing have also been generalized to include time dependence. Indeed, the time-dependent HF (TDHF) is widely used to describe dynamics of many-body systems in nuclear physics and quantum chemistry~\cite{Ring_80,McLachlan_64,Deumens_94}. It is worth noting that the Anderson pseudospin approaches~\cite{Barankov_2004,Barankov_2006,Yuzbashyan_2006a,Yuzbashyan_2006,Gurarie_2009} to quantum quenches of SC or SDW orders are also based on TDHF.

Generalization of the mean-field methods to dynamical evolution of interacting systems can be achieved via the Dirac-Frenkel time-dependent variational principle (TDVP)~\cite{Dirac_1930,Frenkel_1934}. By constraining a quantum state to a specific manifold of the Hilbert space through a variational wave function, the minimum action equation gives an effective dynamical description for the variational parameters. For example, applying TDVP to variational matrix product states (MPS) offers an alternative approach to introduce time evolution in DRMG-based methods~\cite{Haegeman_2011,Haegeman_2016}. The TDHF methods, or more generally the time-dependent self-consistent field methods for fermions, can similarly be derived by constraining the many-body wave function to the form of Slater determinant in the TDVP formalism. Here we outline the approach for a generalized $t$-$V$ model with both particle hopping $t_{ij}$ and density-density interaction $V_{ij}$ extended to further neighbor pairs $(ij)$. The derivation can be straightforwardly generalized to other lattice models. 

We start with the Dirac-Frenkel action for a general quantum state $|\Psi(t)\rangle$ whose dynamics is governed by a Hamiltonian $\hat{\mathcal{H}}$.
\begin{equation}
    S[\Psi] = \int dt \bra{\Psi(t)}\left( i\hbar \frac{d}{dt} - \hat{\mathcal{H}} \right)\ket{\Psi(t)}
\end{equation}
The equation of motion for the quantum state is given by the least action principle: $\partial S / \partial \langle \Psi|  = 0$. Indeed, the minimum action equation simply reproduces the Schr\"odinger equation for an unconstrained wave function. 
Next we assume that the many-fermion quantum state is a Slater determinant, which can be expressed as a filled Fermi sea of self-consistently determined quasiparticles: 
\begin{eqnarray}
	\label{eq:SD}
	|\Psi(t) \rangle = \prod_{\mu} \hat{\gamma}^\dagger_\mu(t) |0\rangle, 
\end{eqnarray}
where $\hat{\gamma}_\mu^\dagger(t)$ is the time-dependent creation operator of a quasiparticle with quantum number $\mu$ and $|0\rangle$ is the vacuum of the fermions. The quasiparticle operators are related to the fermion operators $\hat{c}_{i}^\dagger$ through a time-dependent unitary transformation:
\begin{eqnarray}
	\label{eq:gamma-op}
    \hat{\gamma}^\dagger_\mu(t) = \sum_i \phi^\mu_i(t) \, \hat{c}^\dagger_i.
\end{eqnarray}
As will be shown below, the transformation coefficients $\phi^\mu_{i}$ play the role of effective single-particle wave functions. The independence of the quasiparticles further imposes the orthogonality conditions for the wave functions $\sum_{i} \phi^{\mu \, *}_{i}(t) \phi^\nu_{i}(t) = \delta_{\mu\nu}$, which have to be satisfied at all times. 
Substituting the Slater determinant wave function into the Dirac-Frenkel action and using the Wick's theorem to compute the expectation values, we obtain
\begin{eqnarray}
	& & S[\{\phi\}] =  \int dt \sum_\mu \sum_{ij} \left[ \phi^{\mu \, *}_{i} \left(i\hbar \delta_{ij} \frac{d}{dt} + t_{ij} \right) \phi^\mu_j \right] \\
	& & \qquad -  \int dt \sum_{\mu,\nu} \sum_{ij } V_{ij} \left( |\phi^\mu_{i}|^2 |\phi^\nu_j|^2 - \phi^{\mu *}_{i} \phi^{\nu *}_j \phi^{\mu}_j \phi^{\nu}_{i} \right) \nonumber
\end{eqnarray}
The two terms in the second line above correspond to the familiar Hartree and Fock decoupling, respectively. The least action condition $\partial S / \partial \phi^{\mu*} = 0$ gives the following nonlinear single-particle Schr\"odinger equations
\begin{eqnarray}
    \label{eq:TD-schrodinger}
	& & i \hbar \frac{d \phi^{\mu}_{i}}{dt} = -\sum_j  t_{ij} \phi^{\mu}_j +  \sum_j  \sum_{\nu} V_{ij} |\phi^{\nu}_j|^2 \phi^{\mu}_{i} \nonumber \\
	& & \qquad \qquad -  \sum_j  \sum_{\nu} V_{ij} \phi_j^{\nu *} \phi_{i}^{\nu} \phi^{\mu}_j.
\end{eqnarray}
The coupled differential equations of these single-particle wave functions give a complete dynamical description of the many-fermion systems. Importantly, for a given set of initial wave functions $\phi^{\mu}_{i}(t = 0)$ that are normalized and orthogonal to each other, it can be shown that the orthogonality conditions are maintained by the above dynamical equations.

\section{Anderson pseudospin formulation}
\label{sec:k-space}

Here we derive an effective spin formulation for the coherent dynamics of CDW from the general TDHF equations. For a bipartite lattice, the nearest-neighbor repulsive density-density interaction naturally leads to a disparity of particles on the two sublattices. The resultant long-range order corresponds to a commensurate CDW state with an ultra-short modulation period. In the case of square lattice, this CDW order is characterized by a checkerboard pattern associated with the wave vector $\mathbf Q = (\pi, \pi)$. For a time-dependent CDW state, an effective order parameter for the charge modulation is
\begin{eqnarray}
	\label{eq:CDW-delta1}
	\Delta(t) = \frac{1}{N} \sum_{i} \langle \Psi(t) | \hat{n}_{i} |\Psi(t) \rangle e^{i \mathbf Q \cdot \mathbf r_{i}}, 
\end{eqnarray}
where $\hat{n}_{i} = \hat{c}^\dagger_{i} \hat{c}^{\,}_{i}$ is the number operator of fermions at site-$i$, and the phase factor $\exp(\mathbf Q \cdot \mathbf r_{i}) = \pm 1$ for the two sublattices. Within the TDHF framework, the time-dependent many-fermion state $|\Psi(t)\rangle$ is to be approximated by the Slater determinant state in Eq.~(\ref{eq:SD}). Assuming that the post-quench system remains in a homogeneous CDW state, which means the $\mathbf Q = (\pi, \pi)$ checkerboard pattern is the only Fourier mode of the particle fluctuations, there are only two different values of the on-site particle number depending on the sublattices:
\begin{eqnarray}
\label{eq:bipartite}
	n_{A/B}(t) = \overline{n} \pm \Delta(t).
\end{eqnarray}
Here $\overline{n} = 1/2$ is the average number of particles per site. Importantly, the assumption that the time-evolving post-quench system preserves a homogeneous CDW order indicates an emergent translation symmetry associated with the checkerboard pattern, which is a 45$^\circ$-rotated square lattice with a doubled unit cell.  This in turn implies that wave vectors $\mathbf k$, restricted to the reduced Brillouin zone, are good quantum numbers of the time-dependent CDW states. We thus introduce the following ansatz for the wave functions of quasiparticles
\begin{eqnarray}
	\label{eq:phi-k}
	\phi^{\mathbf k}_{i}(t) = \frac{e^{i \mathbf k \cdot \mathbf r_{i}}}{\sqrt{N}} \eta^{s_{i}}_{\mathbf k}(t),
\end{eqnarray}
where $s_{i} = A, B$ denotes the sublattice of site-$i$. The time evolution of $\eta^s_{\mathbf k}(t)$ is governed by the Fourier transform of the time-dependent Schr\"odinger equation~(\ref{eq:TD-schrodinger}):
\begin{eqnarray}
	i\hbar \frac{d}{dt} \left(\begin{array}{c} \eta^A_{\mathbf k} \\[6pt] \eta^B_{\mathbf k} \end{array} \right)
	= \left(\begin{array}{cc} -V \Delta(t) & \epsilon_{\mathbf k} \\[6pt] \epsilon_{\mathbf k} &  + V \Delta(t) \end{array} \right)
	\left(\begin{array}{c} \eta^A_{\mathbf k} \\[6pt] \eta^B_{\mathbf k} \end{array} \right), \quad
\end{eqnarray}
where $\epsilon_{\mathbf k} = -2t_{\rm nn} (\cos k_x + \cos k_y)$ is the dispersion relation of square-lattice tight-binding model. Since the CDW order is driven by the Hartree term, we have neglected the Fock exchange term for simplicity. We have also dropped the constant diagonal term $V \overline{n}$ which contributes only to an overall phase of the wave functions. Using the ansatz~(\ref{eq:phi-k}) for the Slater determinant in Eq.~(\ref{eq:CDW-delta1}), the time-dependent CDW order parameter is given by
\begin{eqnarray}
	\label{eq:Delta_1}
	\Delta(t) = \sum_{\mathbf k} \left( \left| \eta^A_{\mathbf k}(t) \right|^2 - \left| \eta^B_{\mathbf k}(t) \right|^2 \right)
\end{eqnarray}
The dynamical equation of the CDW state is then reduced to that of a collection of two-level systems, each associated with a wave vector $\mathbf k$.
An intuitive description of a dynamical two-level system is given by the Anderson pseudospin approach~\cite{Anderson_1958}. To this end, we introduce an effective spin  $\mathbf S_{\mathbf k}$ for each wave vector $\mathbf k$, with the following definition:
\begin{eqnarray}
	S^x_{\mathbf k} &=& \eta^{A *}_{\mathbf k} \eta^B_{\mathbf k} + \eta^{B *}_{\mathbf k} \eta^{A}_{\mathbf k}, \nonumber \\
	S^y_{\mathbf k} &=& -i \left(\eta^{A *}_{\mathbf k} \eta^B_{\mathbf k} - \eta^{B *}_{\mathbf k} \eta^{A}_{\mathbf k} \right), \\ 
	S^z_{\mathbf k} &=& \left| \eta^A_{\mathbf k}(t) \right|^2 - \left| \eta^B_{\mathbf k}(t) \right|^2. \nonumber
\end{eqnarray}
Since the $A$ and $B$ sublattices are related by the wave vector $\mathbf Q = (\pi, \pi)$, as will be shown in the next section, each pseudospin $\mathbf S_{\mathbf k}$ thus represents the dynamical degree of freedom for a fundamental particle-hole pair with momenta $\mathbf k$ and $\mathbf k + \mathbf Q$. The dynamics of these effective spins is governed by the Bloch equation~\cite{Bloch_1946},
\begin{eqnarray}
\label{Bloch_dyn}
	\frac{d \mathbf S_{\mathbf k}}{dt} = \mathbf B_{\mathbf k}(t) \times \mathbf S_{\mathbf k},
\end{eqnarray}
which essentially describes the Landau-Lifshitz precession dynamics with a time-dependent magnetic field
\begin{eqnarray}
	\label{eq:B-field}
	\mathbf B_{\mathbf k}(t) = \frac{2}{\hbar}  \left( \epsilon_{\mathbf k}, \, 0, \, -V\Delta(t) \right).
\end{eqnarray}
Note that the Bloch equations for different spins are coupled to each other through the $z$-component of the magnetic field. From the definition Eq.~(\ref{eq:Delta_1}), we have $B^z(t) = -V \sum_{\mathbf k} S^z_{\mathbf k}(t)$, which shows that the magnetic field that drives the spin dynamics originates from individual pseudospins.

\begin{figure}[t]
\includegraphics[width=85mm]{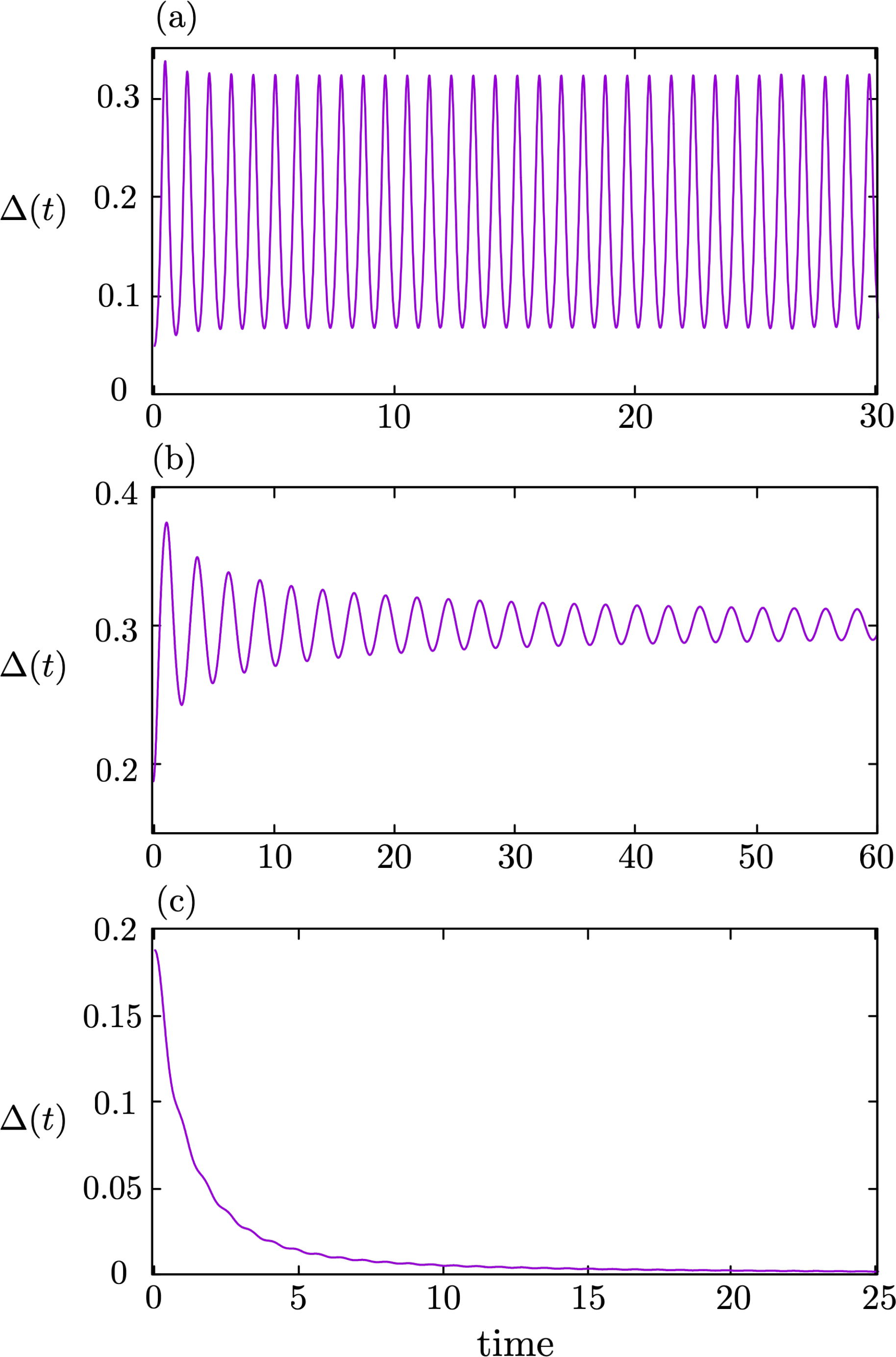}
\caption{\label{CDW_k}
The time dependence of the CDW order parameter $\Delta(t)$ in the three main dynamical regimes of a quantum quench: (a) phase-locked regime with persistent oscillation ($V_{i}=0.2$, $V_{f}=5.0$), (b) Landau-damping regime ($V_{i}=0.5$, $V_{f}=1.0$), and (c) over-damped regime with a dynamical vanishing of the order parameter ($V_{i}=0.5$, $V_{f}=0.1$).}
\end{figure}

In the following, we apply the above coupled Bloch equations to simulate the interaction quench of the CDW order of the $t$-$V$ model at half-filling. As discussed in Sec.~\ref{sec:TDVP}, due to a perfect nesting of the Fermi surface at half-filling, the system is unstable against the formation of checkerboard CDW order for an arbitrarily small $V$. In our interaction quench simulations, the system is initially prepared in the mean-field ground state of the $t$-$V$ model with the interaction fixed at an initial value $V_{i}$. At time~$t = 0$, the interaction is suddenly changed to $V_{f}$. The quantum quench simulation can also be viewed as the time evolution of a system which is initialized in a CDW state stabilized by $V_{i}$ at $t = 0$ and is subject to an time-independent Hamiltonian with~$V_{f}$ for $t > 0$.

Our simulations summarized in Fig.~\ref{CDW_k} show that the post-quench CDW order  exhibits the three major dynamical behaviors common to order parameters of other symmetry-breaking phases discussed in Sec.~\ref{sec:intro}.  First, in the phase-locked regime corresponding to a strong quench with $V_f/V_i \gg 1$, the CDW order parameter exhibits a persistent oscillation after a short transient period, as shown in Fig.~\ref{CDW_k}(a). In this regime, the precessional dynamics of individual pseudospins synchronizes with one another to produce an oscillating $B$-field that drives their own oscillatory dynamics self-consistently. Physically, this describes a synchronized oscillation of the collective CDW order parameter and individual particle-hole pairs. 

% The time dependence of the CDW order parameter is shown in Fig.~\ref{CDW_k} for these three distinct dynamical phases.

For intermediate quenches such that the static CDW order parameters before and after the quench are similar $\Delta_f \sim \Delta_i$, the sudden change of interaction again results in an oscillation of the CDW order parameter.   Yet, the amplitude of the oscillation decreases with time and the CDW order gradually settles to a different value at large time; see Fig.~\ref{CDW_k}(b). More specifically, the decay of the oscillation amplitude follows a $1/\sqrt{t}$ power law, as in the interaction quench of BCS superconductor and SDW.  This dynamical behavior is similar to the Landau-damping phenomena in plasma physics and numerous other physical systems. The decaying oscillation of the order parameter results from a dissipationless energy transfer from the collective modes to quasiparticle excitations~\cite{Barankov_2006,Volkov_1974}. This damping can also be understood as a result of the increasingly incoherent precessional motions of pseudospins which fail to produce a self-sustaining driving force.

Finally, for quenches toward to a much smaller interaction, $V_f \ll V_i$, the post-quench state exhibits a dynamical vanishing of the CDW order as shown in Fig.~\ref{CDW_k}(c). In this regime, the magnetic field in Eq.~(\ref{eq:B-field}) is dominated by the kinetic energy term. As the precession dynamics of individual particle-hole pairs resorts to their respective natural frequencies, the resultant dephasing leads to an overdamping of the CDW order. It is worth noting that here the quenched system relaxes to a state with a vanishing CDW, despite the fact that a finite CDW order is expected to exist for a nonzero $V_f > 0$ due to the Fermi-surface nesting instability discussed above.  Although the system cannot really thermalize under the TDHF evolution, the vanishing CDW order can be intuitively understood as a result of the system being in a quasi-equilibrium state of a temperature that is higher than the critical $T_c$ of $V_f$. Inclusion of dissipative mechanisms in the post-quench evolution is thus expected to bring the CDW order back to the static $\Delta_f$.

We also note that, instead of an exponential decay as in the case of quench dynamics of BCS pairing, the decline of the CDW order, which sometimes is accompanied by a small oscillation, follows a $1/t$ power-law, similar to the case of SDW order in the quantum quench of square lattice Hubbard model~\cite{Blinov_2017}. This algebraic decay could be attributed to the non-analyticity of the density of states of the square-lattice tight-binding model at half-filling~\cite{Blinov_2017}.

\section{Real-space von~Neumann equation}
\label{sec:real-space}

The Anderson pseudospin approaches discussed in the previous section are widely used in the study of quantum quenches of various symmetry-breaking phases including superconductivity and spin-density waves. As the pseudospin methods assume the persistence of a perfect long-range order after the quench, these previous works thus preclude any spatial inhomogeneity in the post-quench states. 
To go beyond the pseudospin methods and allow for spatial fluctuations after a quantum quench, here we discuss an efficient real-space formulation of TDHF in terms of correlation function $\rho_{ij} = \langle \hat{c}^\dagger_j \hat{c}^{\,}_i \rangle$, also known as the single-particle density matrix. From the definition of the Slater determinant in Eqs.~(\ref{eq:SD}) and~(\ref{eq:gamma-op}), the density matrix is related to the time-dependent single-particle wave functions as
\begin{eqnarray}
	\rho_{ij}(t) = \langle \Psi(t) | c^\dagger_j c^{\,}_{i} |\Psi(t) \rangle = \sum_\mu \phi^{\mu}_{i}(t) \phi^{\mu *}_j(t),
\end{eqnarray}
Compared with time-dependent Schr\"odinger equation~(\ref{eq:TD-schrodinger}) for the effective wave functions, the density matrix approach has the advantage that $\rho_{ij}$ is directly related to observables and the corresponding equation of motion is amenable to efficient numerical simulations. 

The pseudospins $\mathbf S_{\mathbf k}$ in Sec.~\ref{sec:k-space} actually correspond to the Fourier transform of the density matrix elements, which are given by
\begin{eqnarray}
	\label{eq:rho_qq}
    \rho_{\mathbf q', \mathbf q} = \bigl\langle \hat{c}^\dagger_{\mathbf q'} \hat{c}^{\,}_{\mathbf q} \bigr\rangle,
\end{eqnarray}
where $\hat{c}_{\mathbf q} = (1/\sqrt{N}) \sum_{i} \hat{c}_{i} e^{-i \mathbf q\cdot \mathbf r_{i}}$
denotes the fermion operator in momentum space. In the presence of a perfect CDW order, the only nonzero matrix elements are $\rho_{\mathbf k, \mathbf k}$ and $\rho_{\mathbf k, \mathbf k \pm \mathbf Q}$, where wave vectors $\mathbf k$ are now restricted to the reduced Brillouin zone. The effective spins are then given by
\begin{eqnarray}
	& & S_{\mathbf k}^x = \frac{1}{2} \left( \rho_{\mathbf k, \mathbf k} - \rho_{\mathbf k+\mathbf Q, \mathbf k + \mathbf Q} \right),  \\
	& & S_{\mathbf k}^y =  -{\rm Im}\,\rho_{\mathbf k, \mathbf k+\mathbf Q}, \qquad
	S_{\mathbf k}^z = {\rm Re} \, \rho_{\mathbf k, \mathbf k +\mathbf Q}. \nonumber
\end{eqnarray}
This simple analysis further highlights the fact that the pseudospin approach precludes inhomogeneous post-quench states. The onset of spatial fluctuations thus corresponds to the emergence of nonzero matrix elements $\rho_{\mathbf k, \mathbf k + \mathbf q}$ with incommensurate wave vectors $\mathbf q \neq \mathbf Q$.

\begin{figure}[t]
\includegraphics[width=85mm]{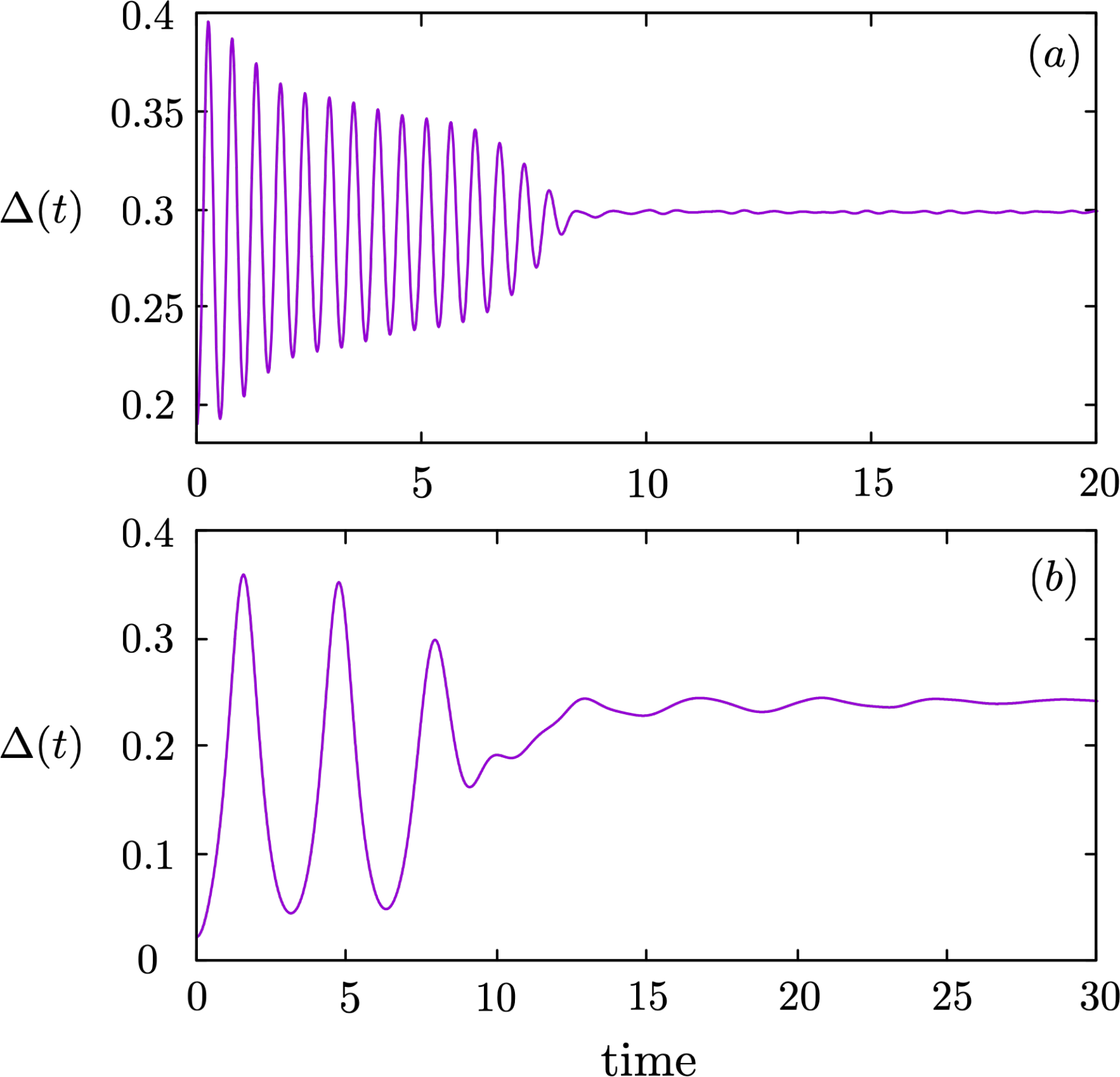}
\caption{\label{inhomo_t}
The overall CDW order parameter $\Delta(t)$ as a function of time in the regime of spatially inhomogeneous post-quench states. The quench parameters are: (a) $V_{i}=0.5$ to $V_{f}=5.0$, and (b) $V_{i}=0.1$ to $V_{f}=1.5 $. Snapshots of the emergent patterns are shown in Fig.~\ref{snapshots1} and~\ref{snapshots2}. }
\end{figure}

\begin{figure*}[t]
\includegraphics[width=175mm]{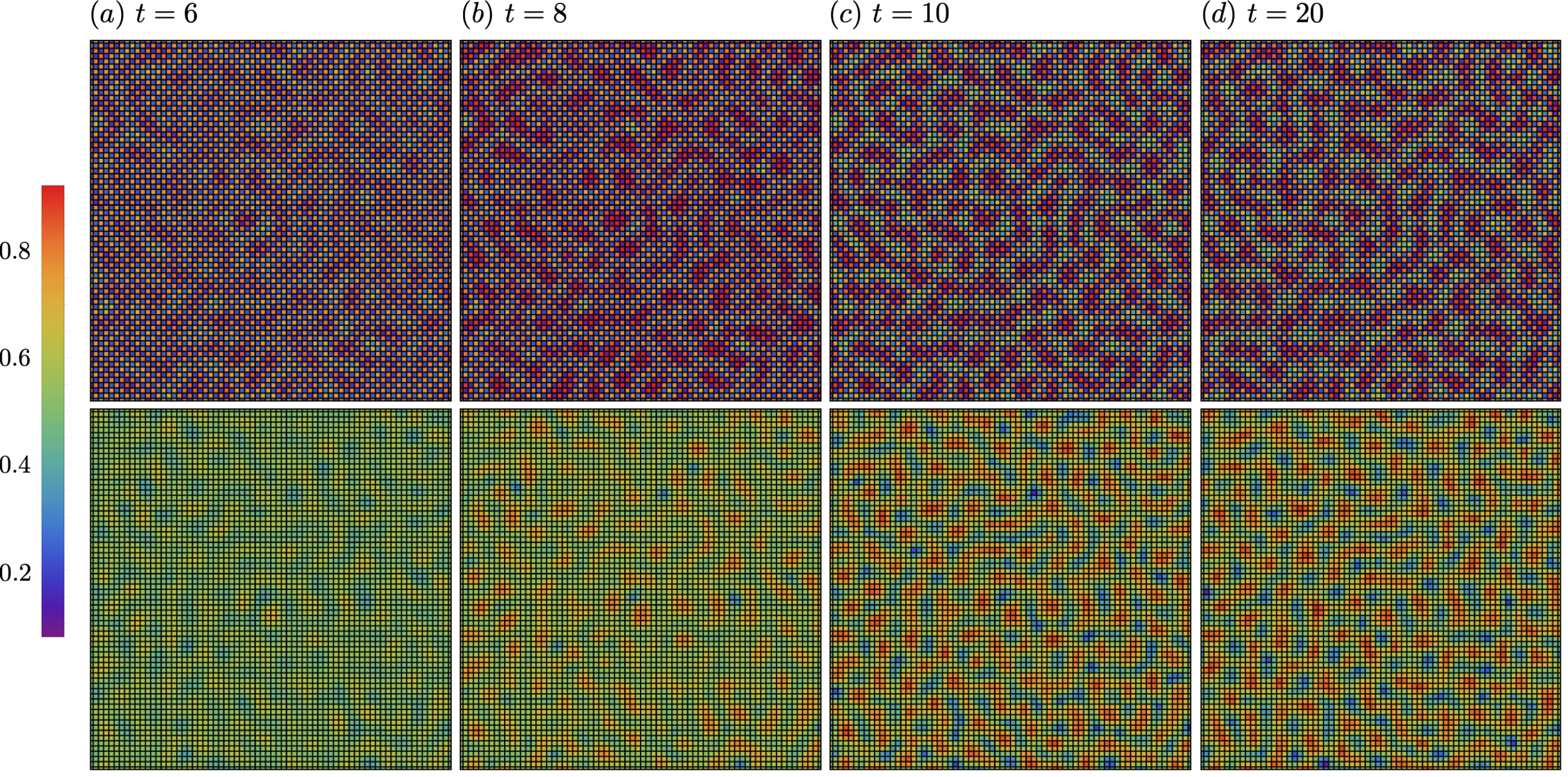}% Here is how to import EPS art
\caption{\label{snapshots1}
Snapshots of particle density $n_{i}=\rho_{ii}$ distributions (top) and local CDW order parameter $\phi_i$ (bottom) at different time steps after a quantum quench from $V_{i}=0.5$ to $V_{f}=5.0$. The corresponding checkerboard CDW order as a function of time is shown in Fig.~\ref{inhomo_t}(a). Note that the Ising order parameter field, while exhibiting complex patterns, remains positive throughout the relaxation process.  }
\end{figure*}

The dynamical equation that governs the time dependence of $\rho_{ij}$ can be directly obtained from the time-dependent Schr\"odinger equation~(\ref{eq:TD-schrodinger}). An alternative, which is physically more intuitive, is to use the ``first-quantized" formulation of the mean-field Hamiltonian. To this end, we note that from the standard mean-field decoupling for the interaction term: $\hat{n}_{i} \hat{n}_j\rightarrow \langle \hat{n}_{i}\rangle \hat{n}_j + \hat{n}_{i}\langle \hat{n}_j\rangle = \rho_{ii} \hat{n}_j + \rho_{jj} \hat{n}_{i}$, one can define a time-depndent mean-field Hamiltonian 
\begin{eqnarray}
    \hat{\mathcal{H}}_{\rm MF}(t) 
    = \sum_{ij} \hat{c}^\dagger_{i} \, H_{ij}[\rho(t)]\, \hat{c}^{\,}_j,
\end{eqnarray}
where the first-quantized Hamiltonian has a simple form
\begin{eqnarray}
    \label{eq:H_{i}j}
    H_{ij} = -t_{ij} + \delta_{ij} \, v_{i}(t).
\end{eqnarray}
The first term represents the kinetic energy, and the second term denotes an effective on-site potential
\begin{eqnarray}
    v_{i}(t) = V \sum_j\!' \rho_{jj}(t)
\end{eqnarray}
where $\sum_j\!'$ indicates summation over sites $j$ that are nearest neighbors to $i$. This Hamiltonian can also be read directly from the form of the time-dependent Schr\"odinger equation in Eq.~(\ref{eq:TD-schrodinger}). The single-particle density matrix then satisfies a self-consistent nonlinear von~Neumann equation, 
\begin{eqnarray}
    -i \hbar\frac{d\rho}{dt} =  [\rho, H(\rho)],
\end{eqnarray}
where $\rho$ and $H$ denote the $N\times N$ matrix of the density matrix and the first-quantized Hamiltonian, respectively. Using Eq.~(\ref{eq:H_{i}j}), the explicit von~Neumann equation for the density matrix is 
\begin{equation}
\label{von_neumann}
    -i \hbar \frac{d\rho_{ij}}{dt} = (v_j - v_{i}) \rho_{ij} 
    + \sum_k \left(t_{ik} \rho_{kj} - \rho_{ik} t_{kj}\right).
\end{equation}
The CDW order parameter in Eq.~(\ref{eq:CDW-delta1}) can be straightforwardly computed from the diagonal matrix elements
\begin{eqnarray}
	\label{eq:Delta_vN}
	\Delta(t) = \frac{1}{N} \sum_i \rho_{ii}(t) e^{i \mathbf Q \cdot \mathbf r_i}.
\end{eqnarray}
It is worth noting that for a system of $N$ sites, there are $N/2$ pseudospins, each indexed by a wave vector $\mathbf k$ in the reduced Brillouin zone. The computational complexity of solving the coupled Bloch equations is thus of order $O(N)$. On the other hand, the number of independent matrix elements of $\rho$ scales as $N^2$ for general Slater determinant states with potential spatial inhomogeneity. Integration of the von~Neumann equation using a naive matrix-matrix multiplication method would lead to a $O(N^3)$ complexity. The von~Neumann equation formulation allows one to take advantage of efficient sparse matrix multiplication algorithms, thus improving the computational efficiency. 
It is worth noting that even though the TDHF approach essentially reduces the many-body problem to an effective single-particle one, additional complexity is required for evolving a density matrix or equivalently a Slater determinant in order to account for the quantum statistics of identical fermions.

Applying the von~Neumann equation method to simulate quantum quenches of the $t$-$V$ model up to $N = 70\times 70$ systems, we first confirm that, for the same system sizes, the real-space approach exactly reproduces the Landau-damped and over-damped oscillations when the $V_i$, $V_f$ parameters are set to those used in Fig.~\ref{CDW_k}. Yet, for most of the strong quench regime where phase-locked oscillation is expected, our large-scale simulations observe a CDW dynamics that is distinctly different from the three dynamical regimes discussed above; two such examples are shown in Fig.~\ref{inhomo_t}. 
In both cases, the interaction is quenched from a small $V_i$ to a much larger~$V_f$. As discussed in Sec.~\ref{sec:TDVP}, the pseudospin simulations for such strong quench regime predicts a persistent oscillation of the CDW order parameter in the post-quench states. Instead, the CDW time traces shown in Fig.~\ref{inhomo_t} exhibit damped oscillations and a finite asymptotic value, similar to the Landau-damping behavior. 

However, contrary to the algebraic $1/\sqrt{t}$ decay characteristic of the Landau-damping regime, here we observe a much faster decay of the oscillation amplitudes. In fact, the decline of the oscillation is even accelerated before the CDW order eventually collapses to a nearly constant value. This seemingly complex dynamical behaviors in time domain are not caused by a complex order parameter as in some previous studies. Instead, as we show in the next section, the collapsed oscillation of the CDW order is due to the emergence of inhomogeneous post-quench states.

\begin{figure*}[t]
\includegraphics[width=175mm]{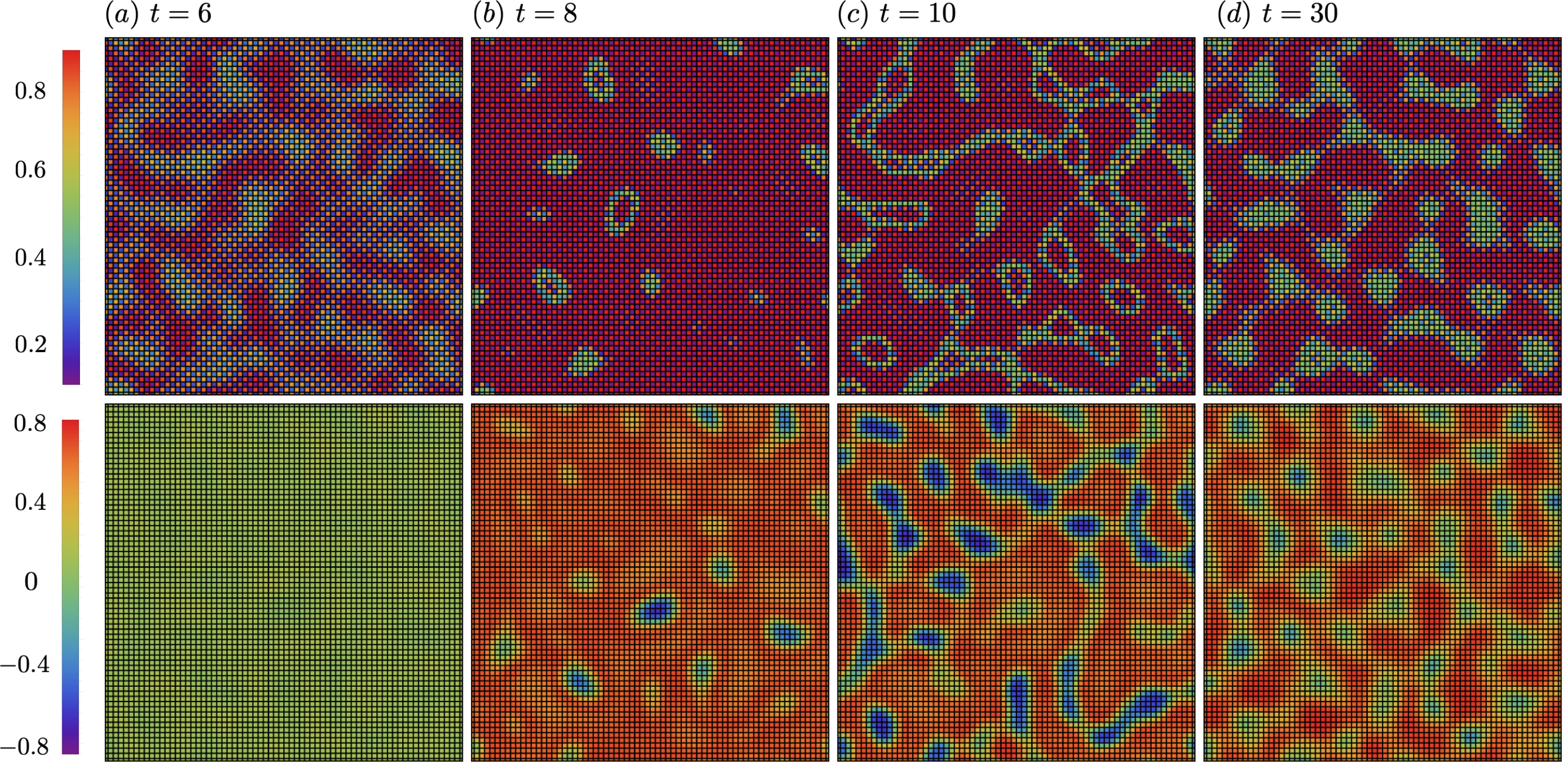}% Here is how to import EPS art
\caption{\label{snapshots2}
Snapshots of particle density $n_{i}=\rho_{ii}$ distributions (top) and local CDW order parameter $\phi_i$ (bottom) at different time steps after a quantum quench from $V_{i}=0.1$ to $V_{f}=1.5$. The corresponding checkerboard CDW order as a function of time is shown in Fig.~\ref{inhomo_t}(b). The interfaces that separate CDW domains of opposite signs correspond to the finite green lines with a nearly constant fermion density of $\rho_{ii} \sim 0.5$ in the top panels.
}
\end{figure*}

\section{Dynamical inhomogeneity and pattern formation}

\label{sec:pattern}

%As discussed in Sec.~\ref{sec:k-space}, the pseudospin approach assumes a perfect checkerboard CDW order throughout the quench process. 

The CDW order parameter $\Delta(t)$ computed in Eq.~(\ref{eq:Delta_vN}) only represents the overall difference of particles in the two sublattices, $\Delta(t) = \bigl(N^e_A(t) - N^e_B(t) \bigr) / N$, yet gives no information about the spatial distribution of particles within the sublattices. 
The unusual damped oscillations in the strong quench regime, as shown in Fig.~\ref{inhomo_t}, thus could be caused by the onset of spatial inhomogeneity. To examine this scenario, the top panels in Fig.~\ref{snapshots1} and~\ref{snapshots2} show the snapshots of on-site particle number $n_i(t) = \rho_{ii}(t)$ at different times after the quench for the two settings in Fig.~\ref{inhomo_t}(a) and (b), respectively. As clearly shown in these snapshots, highly complex patterns develop in the post-quench states.

Careful examinations show that the quench-induced inhomogeneity mostly is in the form of longer wavelength density modulations on top of the checkerboard CDW order. To better characterize such super modulations of particle density, it is convenient to introduce a scalar order parameter field for the local CDW order 
\begin{align}
	\phi(\mathbf r_i) = \Bigl(n_i - \frac{1}{4}\sum_j\phantom{}^{'} n_j \Bigr) \exp\left({i \mathbf Q \cdot \mathbf r_i}\right), 
\end{align}
where the prime in the second term again indicates that the summation is restricted to the nearest neighbors of site-$i$. This local parameter essentially measures the difference of the particle number at a given site and that of its nearest neighbors. A nonzero $\phi_i$ thus indicates the presence of local particle modulation around site-$i$. The phase factor $\exp(i \mathbf Q_i \cdot \mathbf r_i) = \pm1$ is introduced to account for the short-distance checkerboard modulation within a CDW domain. In a perfect checkerboard CDW state, this local order parameter becomes site-independent and is given by the overall CDW order parameter $\Delta(t)$ defined in Eq.~(\ref{eq:Delta_vN}). 

Snapshots of this scalar order parameter $\phi_i$ corresponding to the particle density profiles at the top row of Fig.~\ref{snapshots1} and~\ref{snapshots2} are shown in the respective bottom panels. These results highlight a super-modulation of particle density, as demonstrated by the quasi-periodic square-like patterns in the~$\phi_i$ field, which itself represents a ultrashort-period checkerboard density modulation.

In the first case where the interaction is quenched from $V_i = 0.5$ to $V_f = 5.0$, the local CDW order parameter~$\phi_i$ remains positive throughout the whole system as shown in Fig.~\ref{snapshots1}. This overall positive~$\phi$-field indicates that a long-range coherence of the initial checkerboard CDW is preserved in the strong quench process.  The complex patterns observed in our real-space simulations thus correspond to additional long-wavelength density modulations within a coherent checkerboard domain inherent from the initial CDW state before the quench. This underlying coherent checkerboard order is also consistent with the nearly constant CDW order parameter after the collapse of the oscillations shown in Fig.~\ref{inhomo_t}. If we use the discrete Ising variable $\sigma_i = {\rm sign}\,\phi_i$ to characterize such post-quench states, they would be very different from configurations observed in thermal quenches from a random state to the low-temperature phase of a broken $Z_2$ symmetry~\cite{Bray1994,Puri2009}. Due to the locality of thermal relaxations, a temperature quenches typically results in multiple Ising domains of {\em both} signs $\sigma = \pm 1$ that coexist in a heterogeneous state.

%Fig.~\ref{local-CDW} shows snapshots of this local CDW order parameter at late stages of the post-quench states for quenches from $V_i = 0.5$ to different~$V_f$. Similar to the density plots in Fig.~\ref{snapshots}, complex patterns also develops in the order-parameter $\phi$ field. 

% To characterize the phase coherence of an inhomogeneous CDW state, we further introduce an Ising parameter: $\sigma_i = {\rm sign} \,\phi_i$ which indicates the $Z_2$ nature of CDW domains. 

On the other hand, for quenches starting from a much smaller~$V_i$ as demonstrated in the bottom panels of Fig.~\ref{snapshots2}, Ising domains of opposite signs indeed emerge in the post-quench states. In this case, the initial state before the quench exhibits a homogeneous CDW order with a small positive value $\phi \sim 10^{-2}$. Immediately after the quench, the collapse of the coherent oscillation gives rise to a heterogeneous state where multiple CDW domains with $\sigma_i  = {\rm sign}\,\phi_i = -1$ are embedded in a background of positive CDW order parameter~$\phi$; see Fig.~\ref{snapshots2}(b) and (c). The interfaces that separate Ising domains of opposite signs are of a finite width with a nearly constant particle density of $\rho_{ii} \sim 0.5$. As the system further relaxes, those Ising domains with a negative~$\phi$ gradually revert back to the positive sign, yet with a rather small particle modulation as shown in Fig.~\ref{snapshots2}(d).

\begin{figure}[t]
\includegraphics[width=86mm]{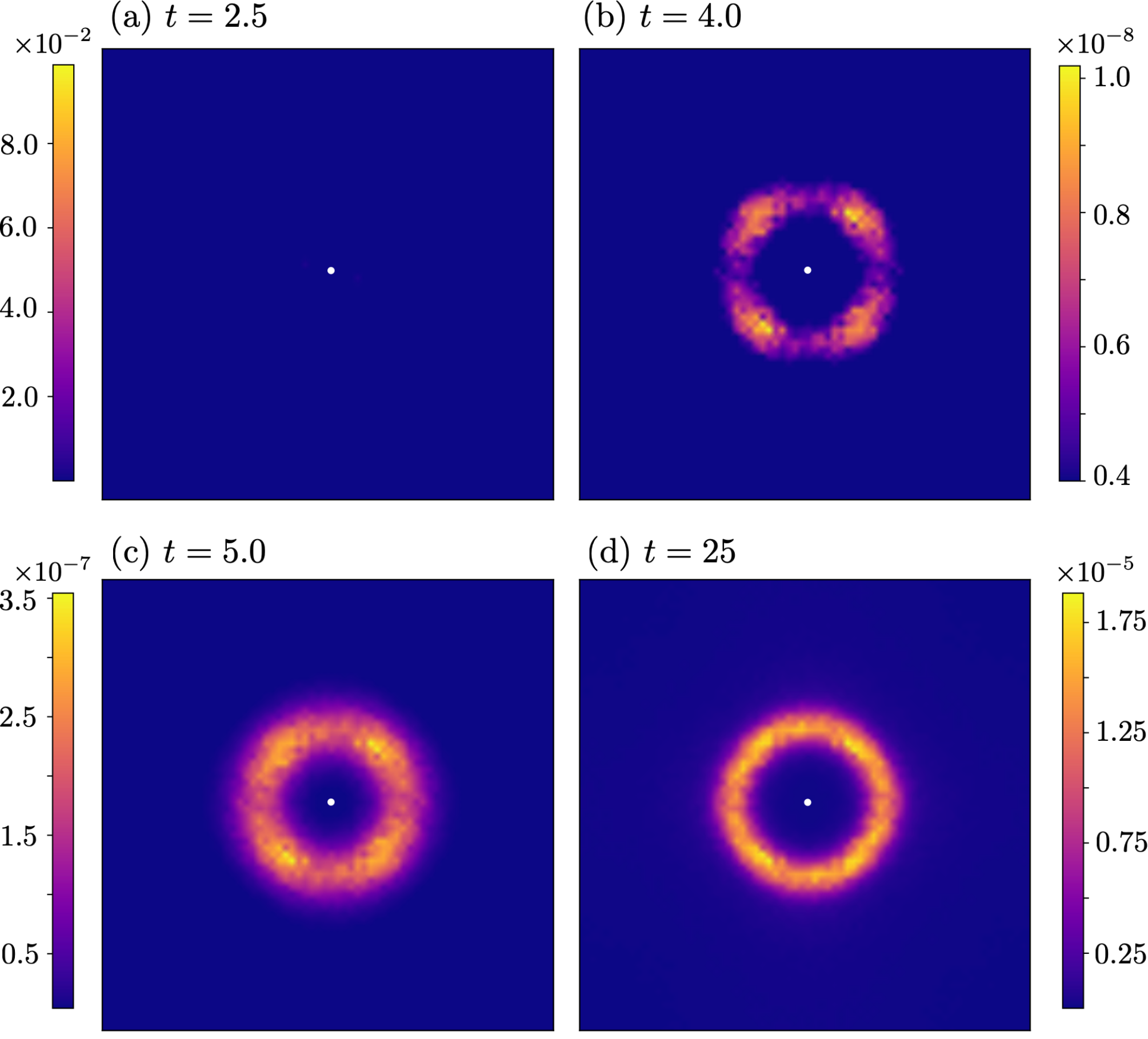}% Here is how to import EPS art
\caption{\label{Sk}
Structure factors $S(\mathbf k, t)$ at various time steps after a quantum quench from $V_{i}=0.5$ to $V_{f}=5.0$ at time $t = 0$. The simulated system size is $N = 70 \times 70$. The results are obtained by averaging over 90 independent von~Neumann dynamics simulations. The white dot at $\mathbf Q = (\pi, \pi)$ corresponds to a dominant checkerboard CDW order. The scale of the color bars in panels~(b)--(d) is chosen to highlight the emergent unstable modes. }
\end{figure}

\begin{figure}[t]
\includegraphics[width=80mm]{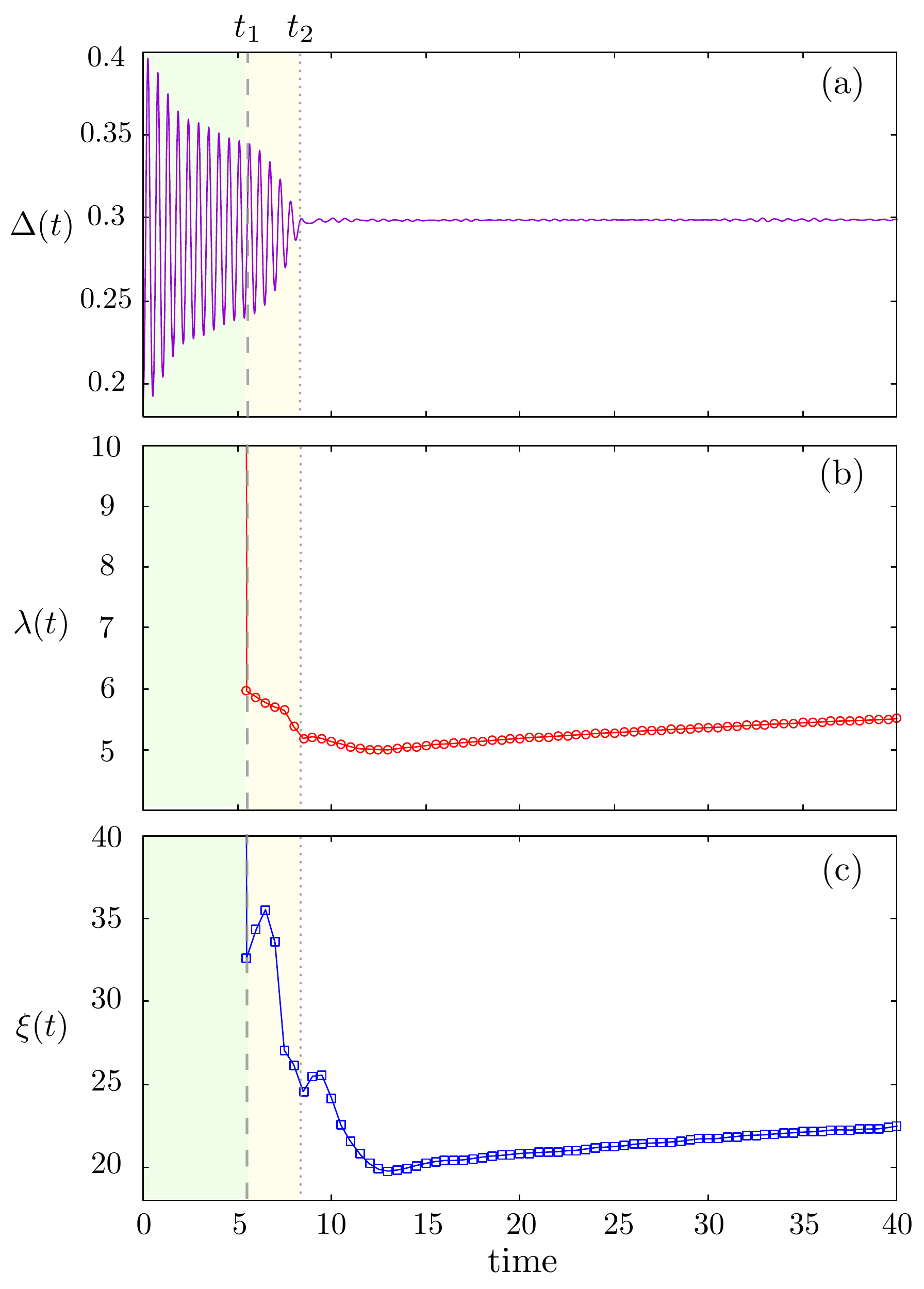}
\caption{\label{traces2}
Post-quench dynamics in the regime with emergent pattern formation for a quench from $V_{i}=0.5$ to $V_{f}=5.0$: (a) time dependence of the CDW order parameter $\Delta(t)$, (b) Modulation length $\lambda(t)$, and (c) correlation length versus time. The two characteristic times $t_1$ and $t_2$ mark the onset of spatial inhomogeneity and the collapse of coherent oscillations, respectively.}
\end{figure}

To further characterize the emergent density patterns, we compute the corresponding structure factor 
\begin{equation}
\label{structure_{f}actor}
	S(\mathbf{q}, t)=|\tilde{n}(\mathbf{q}, t)|^2,
\end{equation}
where $\tilde{n}(\mathbf{q}, t)$ is the Fourier transform of the time-dependent particle density $n_i(t) = \rho_{ii}(t)$. In terms of the momentum-space density matrix introduced in Eq.~(\ref{eq:rho_qq}), the amplitude of density modulation at wave vector $\mathbf q$ is 
\begin{equation}
	\label{ft_e_density}
	\tilde{n}(\mathbf q, t) = \frac{1}{\sqrt{N}} \sum_{\mathbf k} \rho_{\mathbf k, \mathbf k + \mathbf q}(t).
\end{equation}
Fig.~\ref{Sk} shows the structure factors of the post-quench states at a few selected times after a quench from $V_i = 0.5$ to $V_f = 5$. These figures are obtained by averaging over 90 independent von~Neumann dynamics simulations on a $N = 70\times 70$ system. 
In the early stage of the post-quench relaxation, the coherent oscillation of a spatially homogeneous CDW order gives rise to a delta-function peak in the structure factor shown  in Fig.~\ref{Sk}(a). As the system further relaxes, the emergence of spatial inhomogeneity corresponds to the gradual growth of a ring-like feature around the delta-peak, as shown in Figs.~\ref{Sk}(b)--(d). The higher intensity points on the ring correspond to the unstable modes that lead to the pattern formation. The ring-like feature is also characteristic of the labyrinthine structures observed in many pattern-forming systems. Labyrinths are spatial structures where local stripe orders with random orientations fail to develop into a long-range order~\cite{Newell_1993,LeBerre_2002}.

%Indeed, a stripe order is represented by a pair of wave vectors $\pm \mathbf q$, and the lack of long-range orientational order leads to a formation of a ring. 

In our case, the super modulations of particle densities shown in Fig.~\ref{snapshots1} seem to be a combination of local stripe and local checkerboard patterns. A domain of local stripe modulation on top of the ultrashort period checkerboard CDW is represented by a pair of wave vectors $\mathbf Q \pm \eta \hat{\mathbf n}$, where $\eta$ denotes the modulation wave number and $\hat{\mathbf n}$ is a unit vector indicating the orientation of the stripe. Similarly a domain of checkerboard super-modulation on top of the $(\pi, \pi)$ CDW order is characterized by a quadrupole of wave vectors $\mathbf Q \pm \eta \hat{\mathbf n}_1$ and $\mathbf Q \pm \eta \hat{\mathbf n}_2$,  where $\hat{\mathbf n}_1$ and $\hat{\mathbf n}_2$ are a pair of orthogonal unit vectors. The inhomogeneous CDW states obtained in our real-space simulations can be viewed as consisting of multiple domains with random unit vectors $\hat{\mathbf n}$. The lack of long-range orientational order thus leads to the formation of a ring centered at $\mathbf Q$ as shown in Fig.~\ref{Sk}(d).

The modulations of the local CDW order parameter can be characterized by two length scales: the modulation period $\lambda$ of the superstructure (stripes or checkerboard) and the correlation length $\xi$, or characteristic size of superstructure domains. These two length scales are related to the ring feature in the structure factor $S(\mathbf q)$ discussed above. The modulation length is given by the inverse of the radius $\eta$, while the correlation length is proportional to the inverse of the width of the ring. Specifically, the modulation period can be computed as $\lambda = 2\pi / \langle |\mathbf q - \mathbf Q| \rangle$, where the angle bracket indicates average using the structure factor $S(\mathbf q)$ as the weighting factor. A similar formula for the averaged width of the ring can be used to compute the correlation length.

The evolution of these two length scales after a quench from $V_i = 0.5$ to $V_f = 5.0$ is shown in Fig.~\ref{traces2}. Two characteristic times $t_1$ and $t_2$ are introduced to characterize the initial relaxation of the post-quench states. As a homogeneous CDW order is preserved in the initial oscillation period,  the correlation length is of the order of the system size and the modulation period is ill-defined. The onset of inhomogeneity is signaled by a sudden drop of the correlation length at time $t_1$. We also note that the super-modulation patterns corresponding to a finite $\lambda$ emerges immediately at the onset of inhomogeneity at~$t_1$. This indicates that the pattern formation here is generated by a quick growth of the unstable modes that becomes significant at $t_1$.  The emergence of spatial inhomogeneity also accelerates the decline of the oscillation amplitude, and the CDW order $\Delta$ settles to a constant at time $t_2$, although small fluctuations can still be seen afterward.

\begin{figure}[t]
\includegraphics[width=88mm]{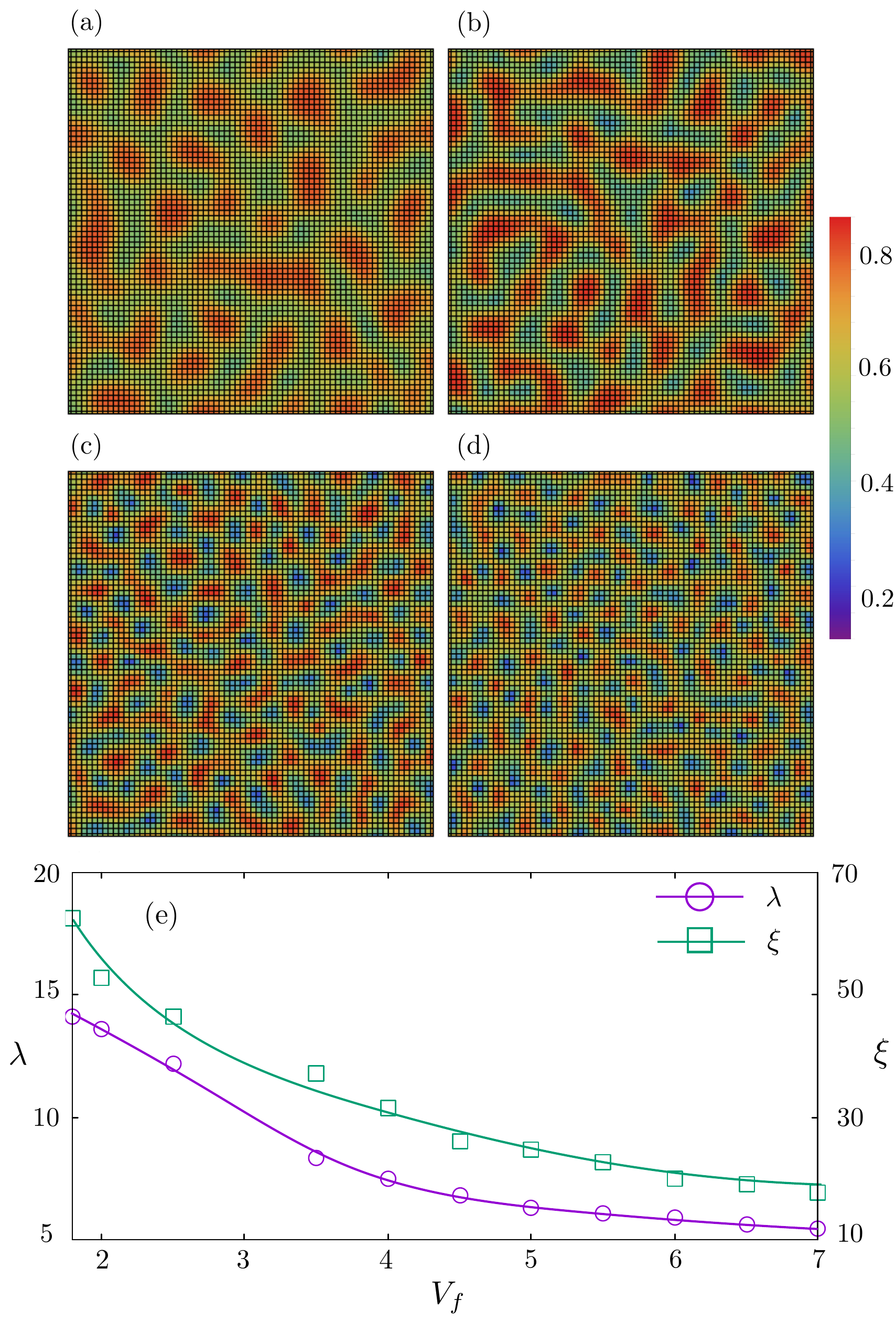}% Here is how to import EPS art
\caption{\label{lengths} 
Local CDW order parameter field at late stage of the quantum quench for (a) $V_{f}=2.0$, (b) $V_{f}=2.5$, (c) $V_{f}=4.0$, and (d) $V_{f}=5.0$ with the initial interaction set at $V_i = 0.5$. 
(e) The modulation length \(\lambda\), and the correlation length \(\xi\) versus $V_f$ with a fixed initial $V_i = 0.5$.}
% These examples indicate that the CDW exhibits spatial inhomogeneity occurs when $V_{f}/V_{i}\sim 3$;
\end{figure}

We also note that both length scales $\lambda$ and $\xi$ of the quasi-steady states at late stage of the relaxation depend on the depth $V_f/V_i$ of the interaction quench. This is demonstrated by the snapshots, shown in Figs.~\ref{lengths}(a)--(d), of the late-stage local CDW order parameter  $\phi_i$ for four different final $V_f$ with the same initial interaction $V_i = 0.5$. Long-wavelength super modulations can be clearly seen in an otherwise overall positive $\phi_i$ field. As discussed above, this overall positive CDW order indicates that a coherent $(\pi, \pi)$ checkerboard density modulation persists in the background for these quench parameters. The modulation period and coherent length of these superstructures versus the final $V_f$ are shown in  Fig.~\ref{lengths}(e) at time $t \sim 200$ after the quench. Both length scales decrease with an increasing~$V_f$. Overall, the post-quench system exhibits a higher level of inhomogeneity with a stronger quench. The dependence of the modulation period on $V_f$ is related to the instability mechanism for the pattern formation, which will be left in future work. On the other hand, the reduced correlation length~$\xi$ in the presence of a stronger quench is likely due to the locality of the pattern formation process in a scenario similar to the Kibble-Zurek mechanism. 

%The decrease of the modulation period $\lambda$ with an increasing $V_f$ indicates unstable modes with larger momenta are induced by a stronger quench. Explanation of this behavior, however, requires an understanding of the instability mechanism, which will be left for a future study.  

\section{Conclusion and Outlook}

\label{sec:conclusion}

To summarize, we have presented a comprehensive study of quantum quench dynamics of CDW states in the $t$-$V$ model.  For the square-lattice model at half-filling, the checkerboard CDW order can be well described by a self-consistent mean-field theory, which can be generalized to include dynamics via the time-dependent variational principle. This approach essentially describes the evolution of a Slater determinant CDW state and is equivalent to the well-known time-dependent Hartree-Fock (TDHF) theory. Assuming that a homogeneous CDW state is maintained throughout the interaction quench, we show that the TDHF is reduced to an Anderson pseudospin theory, which is widely used for the coherent dynamics of order parameter field in the collisionless limit. Our extensive quench simulations produce three dynamical behaviors of the CDW order, namely, the phase-locked persistent oscillation, Landau-damped oscillation, and dynamical vanishing of the order parameter. These are consistent with previously reported dynamical phases induced by interaction quenches of symmetry-breaking phases.

To incorporate spatial inhomogeneity into the quench dynamics, an efficient nonlinear von~Neumann equation is obtained from the real-space TDHF theory to describe the evolution of the post-quench CDW state.  For strong quenches starting from a relatively small initial interaction, our large-scale real-space simulations uncover complex pattern formation in the post-quench CDW states. The onset of spatial inhomogeneity effectively introduces a dephasing mechanism, leading to the collapse of the otherwise phase-locked oscillations of the CDW order parameter. The quench-induced spatial patterns are characterized by domains of super modulations of the particle density in the form of stripes or checkerboards on top of the original ultrashort-period CDW order. However, the lack of orientational coherence between different domains leads to overall disordered patterns similar to the labyrinthine structures observed in many pattern-forming systems. The resultant structure factor is characterized by a ring-like feature centered at the checkerboard wave vector $\mathbf Q = (\pi, \pi)$. The inverse radius and width of the ring correspond to the super modulation period and coherent length, respectively, of the inhomogeneous CDW state.

%We study the post-quench dynamics of CDW in a system with the simplest symmetry breaking, the $Z_{2}$ translation symmetry breaking. In addition to the observations of phase-locked, Landau damping, and overdamped dynamical regimes, consistent with previous studies conducted in reciprocal space in various many-body systems, we also observe the unique spatial inhomogeneities emerged from the CDW orders in real space dynamics. This new dynamical regime, formed with CDW domains separated by Fermi liquid phases, is a novel pattern formation in post-quench dynamics. Therefore, the pattern formation cannot be casted in reciprocal space simulations due to the broken translation symmetry. Importantly, the new regime can only be observed with small $V_{i}$, and the emerged spatial inhomogeneities are different in typical sizes for different sets of $\{V_{i}, V_{f}\}$. Our results indicate that the averaged size of the CDW domains is related to the ratio of $V_{i}$ and $V_{f}$.

%Our study illuminates that the emerging spatial inhomogeneities is a general feature since they can be observed in models with simplest symmetry breaking. However, the mechanism of the emergent inhomogeneity remains an open question. This study sheds more lights on further understanding of the interplay of particle orders, inhomogeneity, and post-quench dynamics, or even topology if we extend to interacting Haldane model.

The emergence of the super modulation patterns on top of the CDW order is caused by unstable modes with wave vectors $\mathbf q = \mathbf Q + (2\pi / \lambda)\hat{\mathbf n}$, where $\lambda$ is the modulation period. A common mechanism for pattern formation is the parametric instability where a pair of unstable modes grow spontaneously from the decay of an initial driver mode through nonlinear interactions. Indeed, the parametric instability has been shown to result in the decay of a uniform oscillating pairing order parameter and the emergence of an inhomogeneous Cooper pair turbulence state.  A possible scenario for pattern formation of the quenched $t$-$V$ model is the decay of the checkerboard CDW order parameter into a pair of such unstable modes $\mathbf q_1$ and $\mathbf q_2$ through the parametric instability mechanism. However, for general modulation period $\lambda$, the momentum is not conserved in this process $\mathbf q_1 + \mathbf q_2 \neq \mathbf Q$. This consideration thus rules out the instability as the direct decay of the phase-locked oscillation of the CDW order.  On the other hand, detailed examinations of the initial relaxation after a quench show that the unstable modes indeed appear in pairs, suggesting a modified version of the parametric instability.  A detailed study of the instability mechanism will be presented elsewhere. 

Our work underscores the importance of dynamical inhomogeneity in quantum quench dynamics of many-body systems. Indeed, here we demonstrate that the post-quench dynamics of a many-fermion system with a simple broken $Z_2$ symmetry is susceptible to pattern-formation instability, even in the collisionless limit. The emergence of complex patterns in this limit is entirely due to the nonlinear dynamics of order parameter fields. Our results indicate that pattern formation is likely to be a generic feature of quantum quenches with more complex order parameters.  It is worth noting that while pattern formation in classical physics is a well-studied subject, quench-induced spatial inhomogeneity in quantum many-body systems is a relatively unexplored area of research. In addition to the nonlinearity originating from the many-body interactions, the quantum fermionic statistics could play a crucial role in the instability mechanisms. For symmetry-breaking phases characterized by complex ordering structures, the associated pattern formation and mechanisms might be closely related to the topological defects of the corresponding order parameter fields. 
 Finally, inclusion of incoherent processes, such as quantum fluctuations, quasiparticle scattering, and energy dissipation, are expected to produce even richer spatiotemporal dynamical behaviors of the post-quench states.

\begin{acknowledgments}
The work was supported by the US Department of Energy Basic Energy Sciences under Contract No. DE-SC0020330. L. Yang acknowledges the support of Jefferson Fellowship. The authors also acknowledge the support of Research Computing at the University of Virginia.
\end{acknowledgments}

\bibliography{ref}
\end{document}